\documentclass[%
 reprint,
superscriptaddress,
 amsmath,amssymb,
 aps,
 prex,
]{revtex4-2}

\usepackage{graphicx}
\usepackage{dcolumn}
\usepackage{bm}
\usepackage[colorlinks=true,citecolor=blue]{hyperref}

\usepackage[dvipsnames]{xcolor}
\definecolor{strawberry}{rgb}{1.0,0.0,0.5}

\newcommand{\err}{\varepsilon}
\newcommand{\Tast}{T^{\ast}}

\usepackage{enumitem}
\begin{document}

\preprint{APS/123-QED}

\title{Biases in Inverse Ising Estimates of Near-Critical Behaviour}

\author{Maximilian B. Kloucek}
\affiliation{H.H. Wills Laboratory, School of Physics, University of Bristol, Bristol, United Kingdom}
\affiliation{Bristol Centre for Functional Nanomaterials, University of Bristol, Bristol, United Kingdom}

\author{Thomas Machon}
\affiliation{H.H. Wills Laboratory, School of Physics, University of Bristol, Bristol, United Kingdom}

\author{Shogo Kajimura}
\affiliation{Faculty of Information and Human Sciences, Kyoto Institute of Technology, Kyoto 606-8585, Japan}

\author{C. Patrick Royall}
\affiliation{Gulliver UMR CNRS 7083, ESPCI Paris, Universit\'{e} PSL, 75005 Paris, France}

\author{Naoki Masuda}
\affiliation{Department of Mathematics, State University of New York at Buffalo, New York 14260-2900, USA }
\affiliation{Computational and Data-Enabled Science and Engineering Program, State University of New York at Buffalo, Buffalo, New York 14260-5030, USA}

\author{Francesco Turci}
\affiliation{H.H. Wills Laboratory, School of Physics, University of Bristol, Bristol, United Kingdom}
\email{mk14423@bristol.ac.uk}

\date{\today}

\begin{abstract}
Inverse Ising inference allows pairwise interactions of complex binary systems to be reconstructed from empirical correlations. Typical estimators used for this inference, such as Pseudo-likelihood maximization (PLM), are biased. Using the Sherrington-Kirkpatrick (SK) model as a benchmark, we show that these biases are large in critical regimes close to phase boundaries, and may alter the qualitative interpretation of the inferred model. In particular, we show that the small-sample bias causes models inferred through PLM to appear closer-to-criticality than one would expect from the data. Data-driven methods to correct this bias are explored and applied to a functional magnetic resonance imaging (fMRI) dataset from neuroscience. Our results indicate that additional care should be taken when attributing criticality to real-world datasets.

\end{abstract}

\maketitle



\section{Introduction} \label{sec:Introduction}

It is often the case that while interactions between individual constituents of complex systems are unknown, their correlations are measurable. Reconstructing the strength of the interactions from these correlations is an inverse problem. Inverse Ising inference, also known as \textit{pairwise} maximum entropy modelling, is an inference technique used to learn the maximum entropy model 
 (MEM) \cite{jaynes1957Phys.Rev.} representing a system of interacting binary variables \cite{aurell2012Phys.Rev.Lett., nguyen2017AdvancesinPhysics} - termed \textit{spins}. Following seminal work on the inference of interactions in retinal neurons \cite{schneidman2006Nature}, the maximum entropy modelling framework has been used in a range of biological settings, from understanding protein interactions \cite{weigt2009ProceedingsoftheNationalAcademyofSciences} to modelling antibody diversity \cite{mora2010ProceedingsoftheNationalAcademyofSciences} and even to analyse the collective behaviour of flocks of birds \cite{bialek2012ProceedingsoftheNationalAcademyofSciences}. In neuroscience, in particular, the inference of pairwise MEMs (i.e. Ising models) from binary data has become common practice and is used both to understand the behaviour of neuronal tissue \cite{schneidman2006Nature, tang2008JournalofNeuroscience, tkacik2014PLoSComputBiol} and to learn functional connectivity networks from coarse-grained functional magnetic resonance imaging (fMRI) studies in humans \cite{watanabe2013NatCommun, watanabe2014Front.Neuroinform., watanabe2014NatCommun, ezaki2020CommunBiol}. This procedure provides insight into the structure of the inferred networks, including their sparsity and the heterogeneity of their couplings \cite{nguyen2017AdvancesinPhysics, lokhov2018Sci.Adv., decelle2014Phys.Rev.Lett.}.

\begin{figure}
\includegraphics{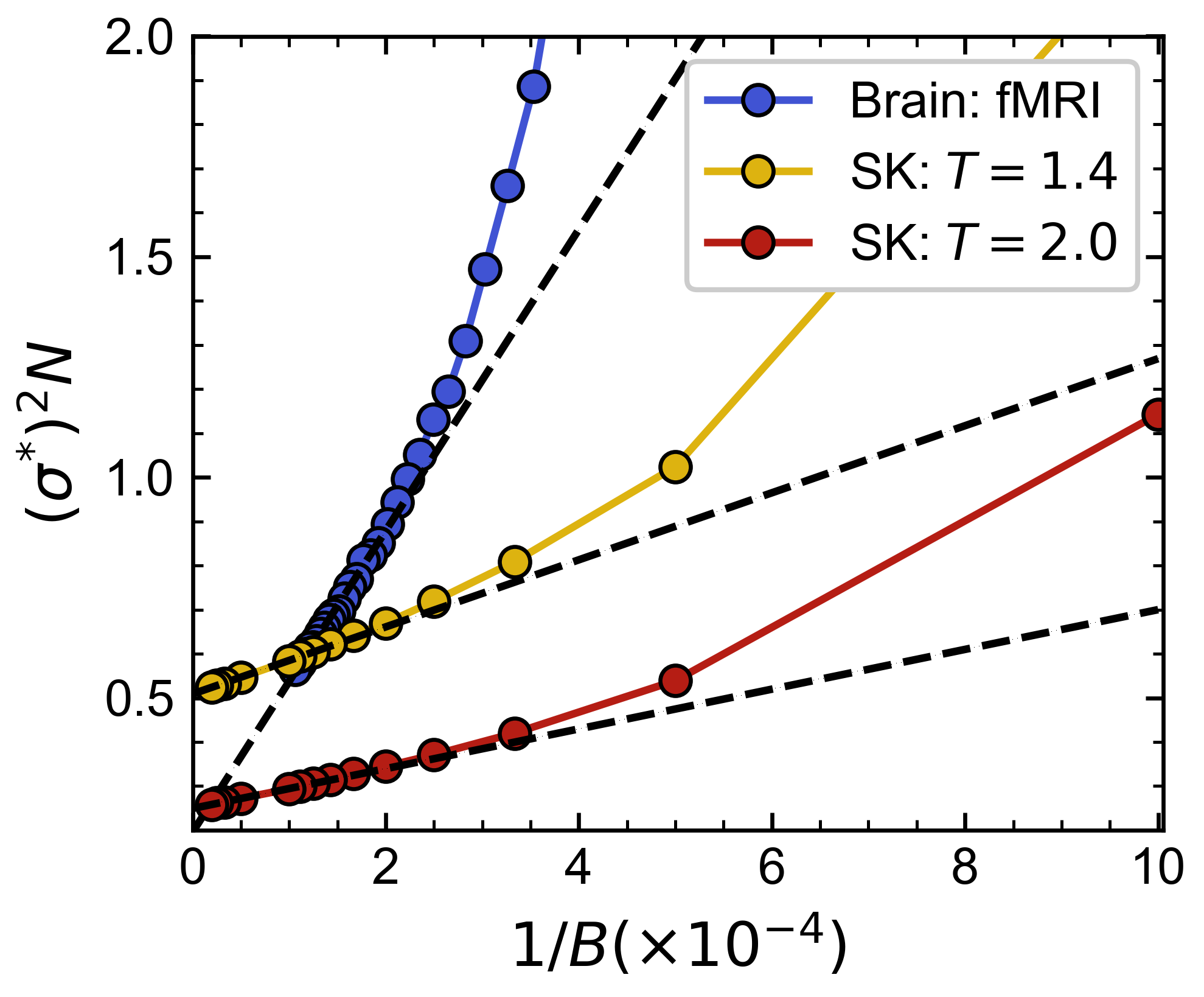}
\caption{Convergence of the variance of the inferred parameters $(\sigma^*)^2$ (scaled by system size) vs inverse sample number $1/B$. The dashed lines indicate linear fits to the asymptotic (large $B$) regime of the data. Asymptotic intercepts of the two SK state-points correspond to $T^{*} = 1.40$ and $T^{*}=2.00$; PLM is able to perfectly reproduce the state-point given infinite data. The gradient $b_1$ of the asymptotic fits sets the severity of the small sample bias. This gradient depends strongly on the state-point, system size and topology of the input data, e.g. we find $b_1(\text{SK: T= 1.4}) = 760$, $b_1(\text{SK: T= 2.0}) = 450$, and $b_1(\text{Brain}) = 3416$.}
\label{fig:bias-overview}
\end{figure}


In such Ising models, critical behaviour emerges between a disordered high-temperature paramagnetic (P) phase with weak correlations and a low-temperature strongly interacting spin-glass (SG) phase with multiple meta-stable minima and large correlations \cite{sherrington1975Phys.Rev.Lett.,kirkpatrick1978Phys.Rev.B, dealmeida1978J.Phys.Math.Gen., parisi1979Phys.Lett., parisi1980J.Phys.Math.Gen., parisi1983Phys.Rev.Lett., castellani2005J.Stat.Mech.}. The critical state is associated with a range of advantageous properties, including providing optimal sensitivity to inputs \cite{kinouchi2006NaturePhys}, enabling coordination between individual elements  \cite{cavagna2010ProceedingsoftheNationalAcademyofSciences, tagliazucchi2012Front.Physiol.}, allowing a large range of dynamic responses \cite{ramo2006JournalofTheoreticalBiology, deco2012JournalofNeuroscience} and maximising computational ability through edge-of-chaos computation \cite{bertschinger2004NeuralComputation, legenstein2007NeuralNetworks}. In neuroscience, the activation patters of ensembles of neurons have been shown to display typical signatures of critical behaviour, as these so-called ``neuronal avalanches'' follow power-law distributions with size and dynamics that are consistent with a critical branching process \cite{beggs2003J.Neurosci., beggs2008PhilosophicalTransactionsoftheRoyalSocietyA:MathematicalPhysicalandEngineeringSciences, haldeman2005Phys.Rev.Lett., beggs2012Front.Physiol., wilting2019CurrentOpinioninNeurobiology}. Similar power-law distributions have also been found in human brain imaging studies \cite{tagliazucchi2012Front.Physiol., priesemann2013PLoSComputBiol}, and inverse Ising inference  specifically has recently shown that typical brain activity measured by fMRI occurs near the SG-P phase transition \cite{ezaki2020CommunBiol}. More generally, a range of evidence supports that many complex biological systems exist in a near critical state \cite{mora2011JStatPhys, munoz2018Rev.Mod.Phys.}, and it is postulated that the advantageous properties of the critical state may generally cause complex systems to organise towards criticality. With this in mind, maximum entropy techniques provide a valuable tool-set with which to further investigate the criticality hypothesis, allowing the inferred models to be assessed in the well-understood framework of equilibrium statistical physics \cite{jaynes1957Phys.Rev.}.



In this paper we focus on the pseudo-likelihood maximisation (PLM) \cite{ravikumar2010Ann.Stat., aurell2012Phys.Rev.Lett.} approach to solving the inverse Ising problem. We chose this logistic regression based method as it is widely acknowledged as the state-of-the-art solution to the inverse Ising problem \cite{decelle2014Phys.Rev.Lett., lokhov2018Sci.Adv., nguyen2017AdvancesinPhysics}. In section~\ref{sec:plm} we introduce 
 the PLM method and highlight alternative inverse Ising solvers. In section~\ref{sec:PLMperformance} we show that parameters estimated via PLM are biased and that statistical averages of the parameters, such as the variance, converge to the true value as a linear function of $1/B$, see Fig.~\ref{fig:bias-overview}. We relate this convergence to the standard small-sample bias of maximum likelihood estimators (MLEs). Using the Sherrington-Kirkpatrick (SK) \cite{sherrington1975Phys.Rev.Lett., kirkpatrick1978Phys.Rev.B} spin-glass model as a benchmark, we find that the small $B$ bias causes the inferred model to appear tuned towards criticality; models inferred using PLM show both a lower temperature and enhanced critical fluctuations than the input model the data was generated from. Similar to previous authors \cite{lokhov2018Sci.Adv., nguyen2017AdvancesinPhysics}, we find that the rate at which the bias is dissipated is state-point (i.e. temperature) dependent, and are the first - to our knowledge - to link the failure of PLM at low temperatures (i.e. for highly correlated data) to the separation effect observed in logistic regression \cite{albert1984Biometrika, zorn2005Polit.Anal.}. In section \ref{sec:correction} we present two corrective procedures to remove the bias, a self-consistency correction and Firth's penalized logistic regression \cite{firth1993Biometrika, heinze2002Stat.Med., zorn2005Polit.Anal.}, and compare their performance by measuring how well they capture the temperature and critical fluctuations of the input dataset. In section \ref{sec:neuro-case-study} we explore the repercussions of the small sample bias in interpreting inference results from real data by considering a fMRI dataset of brain activity related to meditation. Our results lead us to caution against claims of criticality in PLM models, as we show that small sample size biases tune models inferred from dynamical (i.e. fluctuating) data to a closer-to-critical state.

\section{Background: inverse Ising inference via pseudo-likelihood maximisation}
\label{sec:plm}

We will consider systems of $N$ interacting binary variables $s_{i} \in \pm 1$, $i = 1, \ldots N$ which we refer to as \textit{spins}. These spins may represent any binary quantities, such as the magnetic moments of spins in a metal (up, down), or the state of a neuron or region of interest (ROI) in the brain (on, off). The spins fluctuate in time, and for each of the $N$ labelled regions, we have time series of length $B$. The state of the entire spin vector at a time $t'$, $\boldsymbol{s}(t = t')$, is called a \textit{configuration}, and the full dataset of $B \times N$ observations will either be referred to as a \textit{trajectory} or as the \textit{dataset}. Inverse Ising inference corresponds to learning a set of interaction parameters that is likely to reproduce the observations. Configurations of the inferred model will follow the maximum entropy probability distribution \cite{jaynes1957Phys.Rev.}, and by measuring the per-spin magnetisation $m_i =\langle s_i\rangle$ and cross-correlations 

\begin{equation}
    C_{ij} = \langle s_i s_j\rangle -\langle s_i \rangle \langle s_j\rangle, 
\end{equation}
where $\langle\cdot\rangle$ indicate time averages, one infers the so called \textit{pairwise} maximum entropy model. This model is defined by the Hamiltonian

\begin{equation}
\label{eq:skham}
 \mathcal{H}= - \sum_{i} h_i s_{i} - \frac{1}{2} \sum_{i \neq j} J_{ij} s_{i} s_{j},
\end{equation}
where the $J_{ij}$ represent pair-wise couplings between the spins and the $h_i$ are external fields, with the summation index $i \neq j$ running over all non-matching pairs of $i$ and $j$. The inference problem therefore consists in determining the symmetric matrix of couplings $\bm{J}$ (diagonal entries are zero as there are no self-couplings) and vector of fields $\bm{h}$ from the correlations $\bm{C}$ and averages $\bm{m}$. The probability of observing a given configuration $\boldsymbol{s}$ follows the maximum entropy (Boltzmann) distribution:

\begin{equation}
    P(\boldsymbol{s})= \dfrac{1}{Z}\exp{[-\beta \mathcal{H}(\bm{s})]},
\label{eq:boltzmann}
\end{equation}
with $\beta=1/T$ being the inverse temperature and $Z$ the partition function, a normalisation constant. The log-likelihood of observing a given trajectory $\{\boldsymbol{s}\}_{B}$ from the couplings and fields is

\begin{align}
    \label{eq:likelihood}
    \begin{split}
    \mathcal{L}(\boldsymbol{h}, \boldsymbol{J}|\{\boldsymbol{s}\}_{B})& =\beta \sum_{i} h_{i} m_{i}+\\&+ \frac{\beta}{2} \sum_{i \neq j} J_{i j}\left(m_{i} m_{j}+C_{ij}\right)-\log Z.
    \end{split}
\end{align}

The set of parameters $\{ \bm{h} ^{*}, \bm{J}^{*} \}$ which maximises eq.~\ref{eq:likelihood} is the maximum likelihood solution to the inverse Ising problem. When the number of spins $N$ is very small (typically a few tens), it is computationally feasible to perform this optimization directly. However, the problem becomes rapidly intractable with increasing $N$ (the number of possible configurations scales as $2^{N}$), and a range of alternative methods have been proposed to perform the maximisation. This includes Boltzmann learning \cite{ackley1985Cogn.Sci.} which uses Monte Carlo (MC) simulations \cite{metropolis1953TheJournalofChemicalPhysics, hastings1970Biometrika} to evaluate the gradients of (\ref{eq:likelihood}). While its technically possible to compute these gradients with unbounded accuracy, the random nature and computational intensity of MC sampling means this process is also limited to small ($N \sim 120$ \cite{tkacik2014PLoSComputBiol}) system sizes. Various analytical solutions have also been introduced as alternatives, see \cite{roudi2009Front.Comput.Neurosci., roudi2009Phys.Rev.E, nguyen2017AdvancesinPhysics} for reviews, yet most of these require additional assumptions to be made about the system, and often fail in the low temperature (strong coupling) regime, providing more error prone solutions \cite{nguyen2017AdvancesinPhysics}. A powerful alternative approach, and the method studied here, is pseudo-likelihood maximisation (PLM) \cite{aurell2012Phys.Rev.Lett.}. In PLM one replaces the log-likelihood (\ref{eq:likelihood}) by a set of $N$ pseudo-log-likelihoods

\begin{align}
\begin{split}
\mathcal{L}_{r} (h_r, \boldsymbol{J}_r | \{\boldsymbol{s}\}_{B}) &= \frac{1}{B} \sum ^{B} _{t = 1} \ln P_{\{ h_r, \boldsymbol{J}_r \}}(s_{r}(t) | \boldsymbol{s}_{\setminus r}(t)),
\end{split}
\label{eq:PLM}
\end{align}
which depend only on the parameter $h_r$ and the $r^{th}$ row of entries $\boldsymbol{J}_r = \{ J_{rj} \}_{j \neq r}$ to the coupling matrix. We also introduce the conditional probability distribution

\begin{equation}
    P_{\{ h_r, \boldsymbol{J}_r \}}(s_{r} | \boldsymbol{s}_{\setminus r} )= 
1 / (1 + e^{-2 s_{r} [h_{r} + \sum _{r \neq j} J_{rj} s_{j}]})
,
\end{equation}
corresponding to the probability of observing the $r^{th}$ spin in state $+1$ or $-1$ given all other $N-1$ spins. We note that each of the $\mathcal{L}_{r}$ can be maximised independently for each spin, making the problem highly suitable for parallelisation, and that in the limit of $B \to \infty$ the PLM approach to inverse Ising inference is exact. Moreover, the structure of the pseudo-likelihood means that each PLM optimisation is formally identical to logistic regression for which efficient computational algorithms exist. For this work, we perform the regressions using the  \textit{sklearn.linear\_model.LogisticRegression} classifier from the \textit{Scikit-learn} \cite{pedregosa2011J.Mach.Learn.Res.} Python package. Note that the coupling matrix inferred this way is not symmetric, and we therefore always perform a post-inference symmetrising step, setting $\boldsymbol{J}^{*} = \frac{1}{2} [\boldsymbol{J}_{PLM}^{*} + (\boldsymbol{J}_{PLM}^{*})^{T}]$ where $T$ is the transpose of the matrix. $\bm{J}$ can also be interpreted as the weighted adjacency matrix of a complex network \cite{albert2002Rev.Mod.Phys., newman2003SIAMRev.}, which encodes the topology of the model. As such, $l1$-regularized sparsity promoting versions of PLM \cite{ravikumar2010Ann.Stat.} are commonly used for network reconstruction when a large set of parameters are known (or assumed to be zero), and most extensions to PLM have focused on improving its performance for this purpose \cite{decelle2014Phys.Rev.Lett., lokhov2018Sci.Adv.}. When sparsity cannot be assumed un-regularised PLM still offers superior performance to other approximate inverse Ising solvers \cite{nguyen2017AdvancesinPhysics}, and in this work we focus exclusively PLM without regularisation.

\section{PLM accuracy near criticality}
\label{sec:PLMperformance}


Previous work found  that the accuracy of PLM depends on the temperature (or coupling strength) and topology of the underlying true model \cite{lokhov2018Sci.Adv., montanari2009Adv.NeuralInf.Process.Syst., aurell2012Phys.Rev.Lett., nguyen2017AdvancesinPhysics}, and that errors are minimized near the critical point. It should be noted however, that system sizes of only $N$ from $16$ to $64$ were analysed in these studies, and so the critical point is poorly defined. These results are commonly attributed to the theoretical finding that the generalised susceptibility of the Ising model can be related to the Fisher information matrix \cite{mastromatteo2011J.Stat.Mech.}, and that the maximisation of these quantities at the critical point corresponds to a high density of distinguishable models.

While the above statement is certainly true given infinite data, in real applications small sample size biases can dominate the inference. The parameter estimates obtained through PLM, like those of any maximum likelihood estimator (MLE), are well known to depend both on the number of samples $B$ (with a slow $1/B$ convergence) and a prefactor set by the true parameter \cite{firth1993Biometrika} which is not known a priori, so that, for example, 

\begin{equation}
    J_{ij}^{\ast} =  J_{ij}^{0} + \dfrac{ b_{1,ij}({\boldsymbol{h}}^{0}, {\boldsymbol{J} }^0)}{B} + O(B^{-2}).
\label{eq:MLE-bias}
\end{equation}

Here, the superscripts $\ast$ and $0$ denote the inferred and true values respectively, while  $b_{1,ij}({\boldsymbol{h}}^{0}, {\boldsymbol{J}}^{0})$ is the state-dependent prefactor to the leading $1/B$ bias. Collectively, the prefactors $b_{1,ij}({\boldsymbol{h}}^{0}, {\boldsymbol{J}}^{0})$ set the difficulty of learning a given state-point (i.e. model) by increasing or decreasing the amount of data required to dissipate the bias. When expressed this way, we expect that due to the maximised Fisher information at criticality \cite{mastromatteo2011J.Stat.Mech.}, systems near the critical point would have a smaller first order prefactor making them easier to learn. It is also possible to express averaged quantities of the inferred parameters, such as the standard deviation $\sigma$ of the couplings $\boldsymbol{J}$ in terms of the bias, so that

\begin{equation}
\sigma^{\ast}=\sigma^{0} +\dfrac{b_1({\boldsymbol{h}}^{0}, {\boldsymbol{J} }^0)}{B}+ O(B^{-2}),
\end{equation}
where $b_1$ is the combination of the several bias terms on the individual $J_{ij}$ couplings. Again, $b_1$ is an input state dependent prefactor which sets the bias contribution to the inferred standard deviation. We show this relation holds for two select SK state-points, as well as for a real world dataset in Fig.~\ref{fig:bias-overview}.

The above bias expressions are general, and hold for any MLE. But PLM specifically involves performing a set of binary logistic regressions, and these are known to be affected by an additional small sample size issue termed separation \cite{albert1984Biometrika}. Separation occurs when a subset of covariates (e.g. $\boldsymbol{s}_{sep}  \subset \boldsymbol{s}_{\setminus r}$) in the logistic regression can perfectly predict the outcome variable ($s_{r}$), and leads to (theoretically infinitely) large  estimates for the corresponding parameters. Most commonly the separation will be \textit{quasi-complete} and only parameter estimates associated with $\boldsymbol{s}_{sep}$ will be infinite, with the remaining parameter estimates remaining relatively unaffected. In real settings, where the logistic regression is solved numerically, the precise values of the separated parameters will not be infinite and instead depend on the convergence criteria of the numeric optimization scheme \cite{zorn2005Polit.Anal.}. Methods which implicitly remove the first order bias term through modifying the log-likelihood function \cite{firth1993Biometrika} have been shown to control separation \cite{heinze2002Stat.Med., zorn2005Polit.Anal.}.

 
We perform a study of the first order bias of PLM, and re-contextualise the findings of previous authors in terms of these results. In particular, we connect the effects of separation in the datasets (also observed in \cite{aurell2012Phys.Rev.Lett.}) to criticality. For any SK-like model, the highly correlated nature of data drawn either from the critical point or the low temperature SG and F phases means that there will always be a $B$ dependent threshold temperature below which separation will occur, and at which PLM can no longer estimate the parameters correctly. For non-separated data, our work additionally shows that the bias behaves as

\begin{equation}
    b_{1,ij}({\bf J}^0, {\bf h}) \approx  b_{1,ij}(\sigma_0),
\end{equation}
that is, the bias is, to a good approximation, purely a function of the variance of the parameters (i.e. the inverse temperature).

\subsection{Inference of SK models}

 We benchmark the PLM method on a zero-field ($\bm{h}^{0} = 0$), fully-connected Sherrington-Kirkpatrick (SK) model \cite{sherrington1975Phys.Rev.Lett.} with system sizes comparable to typical coarse grained fMRI brain region analyses \cite{ezaki2020CommunBiol, kajimura2020SciRep, glasser2016Nature}, $N=50,100,200,400,800$, collecting a total of $B_{\rm max}=10\ 000$ samples per state-point.

We generate our input couplings $\bm{J} ^{0}$ by drawing from a Gaussian distribution with mean $\mu^{0} = \mu / TN$ and standard deviation $\sigma^{0} = \sigma / T N^{1/2}$. Note that we have absorbed the temperature of eq.~\ref{eq:boltzmann} into our definition of the model parameters. We do this as for real applications the ``temperature'' (in a statistical physics sense) of the system is undefined, and only coupling strengths can be extracted. $\mu$ and $\sigma$ are intensive variables and the state of the system is characterised by the dimensionless average coupling strength $\mu / \sigma$ and temperature  $T /\sigma$. We fix $\sigma = 1$ and sample form the range $\mu \in[0,2]$, $T \in[0.5,2]$, where in the $N \to \infty$ thermodynamic limit, the system explores all of its phases. These phases are a low temperature disordered spin-glass (SG), low temperature ordered ferromagnetic (F) and high temperature disordered paramagnetic (P) phase. From previous findings, we expect the inference to perform best near the phase transitions \cite{aurell2012Phys.Rev.Lett., nguyen2017AdvancesinPhysics}. We produce input time-series for every state point via standard Monte-Carlo simulations \cite{metropolis1953TheJournalofChemicalPhysics, hastings1970Biometrika}, sampling data every $1000 N$ steps. We monitor the autocorrelation time $\tau$ to ensure that subsequent samples are decorrelated and can be considered as independent and identically distributed (i.i.d.). From the series, we compute the spin-glass order parameter (also known as the \textit{overlap})

 \begin{equation}
q = \frac{1}{N} \sum_{i=1} ^{N} \langle s_{i} \rangle ^{2},
\label{eq:Q}
\end{equation}
and the covariance as:
\begin{equation}
C^{2} = \frac{1}{N} \sum_{i , j = 1} ^{N} C_{ij}^{2},
\label{eq:C2}
\end{equation}
which is related to the spin-glass susceptibility $\chi_{SG}=C^2/T^2$ \cite{ezaki2020CommunBiol, fischer1991}. When approaching the SG phase from the higher temperature P phase, the susceptibility (and hence the covariance) increases rapidly, reflecting the development of spontaneous large correlations near the 
 phase transition, i.e. the critical point. We then perform un-regularised PLM inference on each trajectory, as the models are not sparse. The quality of the inference is assessed by measuring the error

\begin{equation}
\err = \sqrt{
\frac{\sum_{i \le j} \left( \theta_{ij} ^{*} - \theta_{ij} ^{0} \right) ^ {2} }
{\sum_{i \le j} \left( \theta_{ij} ^{0} \right) ^ {2} }
},
\label{eq:ErrNguyen}
\end{equation}
a robust aggregate measure of the deviations in parameter estimation previously defined in \cite{nguyen2017AdvancesinPhysics}. Here $\bm{\theta}$ is a symmetric matrix containing all PLM parameters, with $\theta_{ii} = h_i$ and $\theta_{ij} = J_{ij}$ as all $J_{ii} = 0$ (there are no self-couplings).  Note that as the number of couplings $N_{J} = N(N-1)/2$ is much larger than the number of fields $N_{h} = N$, the error is dominated by contributions from the couplings. The biases enter the numerator of this expression and a $1/B$ scaling of the error is expected.

Each generated model will deviate by a small amount from $T$, and so we define the \textit{measured} temperature of each model realisation as $T^{0} = 1/ ( \sigma^{0} N^{1/2} )$, where $\sigma^{0}$ is the standard deviation computed from the realised couplings. We similarly define the measured inferred temperature as $T^{\ast} = 1/ (\sigma^{\ast} N^{1/2})$ where $\sigma^{\ast}$ is the standard deviation of the inferred couplings. This allows us to define a second global metric on the inference quality, i.e. how well the inferred model reproduces the temperature.

\begin{figure*}
\includegraphics[width=\textwidth]{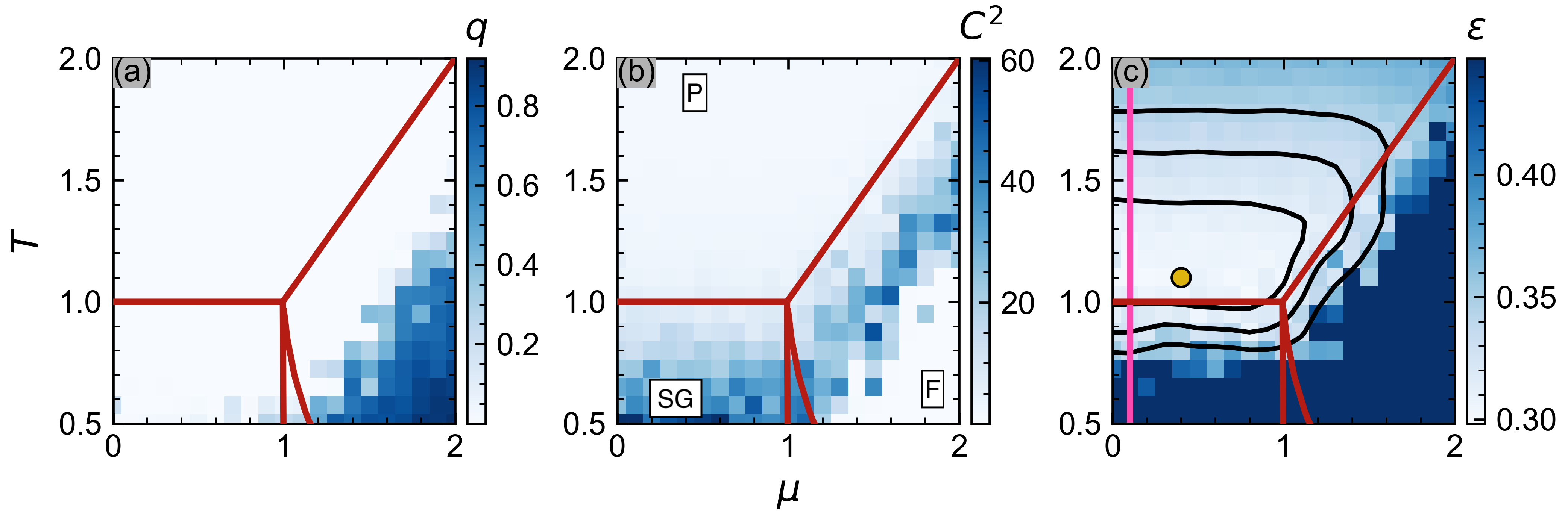}
\caption{
Overview of the order parameter (a), susceptibility (b), and error (c) for the zero-field SK phase diagram with $N=200$ and $B=1 \times 10^4$. The values of the observables shown at each sate-point were calculated by averaging over 3 independent realisations at that state-point. Red lines show phase transition lines in the $N \to \infty$ limit. (b) Labels P, F, and SG indicate locations of paramagnetic, ferromagnetic and spin-glass phases in the thermodynamic limit. (c) Contours of $\err$ are shown in black, with the pink line labelling the line $\mu = 0.1$ across which a more detailed examination of the error is made in Fig. \ref{fig:ObsMuCut}. The location of the minimum error is denoted by a orange circle, and occurs in the P phase above the P-SG boundary. Note that $\err$ is thresholded so that the maximum plotted value is $\err _{\text{max}} = 1.5 \err _{\text{min}}$.
}
\label{fig:SKPD}
\end{figure*}

\subsection{Error dependence on state-point}
\label{sec:SKPD-overview}

In Figure \ref{fig:SKPD} we illustrate the overall phase behaviour of the SK model and of the inference error for $N=200$ and $B=10\ 000$, with the phase boundaries of the $N=\infty$ system overlaid. In (a) we recover the known SK phase diagram, with low values for the overlap in the paramagnetic and spin-glass phase compared to the ferromagnetic phase. The phases transitions correspond to regimes of increased susceptibility, which peak at the phase boundaries but are blurred and shifted due to finite size effects \cite{binder1987Ferroelectrics} as shown in (b). Deep in the F and P phases $C^2$ is low due to the lack of fluctuations in the F phase, and a lack of  spin-spin correlations in the P phase.

Fig. \ref{fig:SKPD}(c) shows the performance of the PLM method as quantified by the error in (\ref{eq:ErrNguyen}). We observe a minimum in $\varepsilon$ in the paramagnetic phase ($\mu=0.4$, $T=1.1$), and a rapid increase as the two correlated F and SG phases are approached. This is consistent with previous studies of the fully-connected SK model for $N=64$ \cite{nguyen2017AdvancesinPhysics, aurell2012Phys.Rev.Lett.}, who found the error to be minimized around $T \sim 1$ for $\mu = 0$. We note importantly that although the error minimum is close to the peak of the critical fluctuations, the two are not coincident - in our simulations we find e.g. the P-SG $C^2$ peak at $T \approx 0.6$, while finite size studies of the SK model have shown the critical fluctuations of the specific heat of the P-SG transition to peak at $T \approx 0.7$ for similar system sizes \cite{aspelmeier2008J.Phys.A:Math.Theor.}. The $\varepsilon$ colour-map is capped at  $1.5 \err_{\text{min}}$ but much larger errors, $\mathcal{O}(10^3 - 10^4) \times \err_{\text{min}}$, are observed in the correlated F and SG phases. These large errors are due to the $B$ configurations becoming highly correlated ($\tau \gg 1$), causing the separation effect introduced in section \ref{sec:PLMperformance} to lead to (infinitely) large parameter estimates. The numeric values of $\varepsilon$ in these regions do not have a real meanings - they simply indicate that no maximum likelihood solution can be found for some of the PLM logistic regressions.

\begin{figure}
\includegraphics{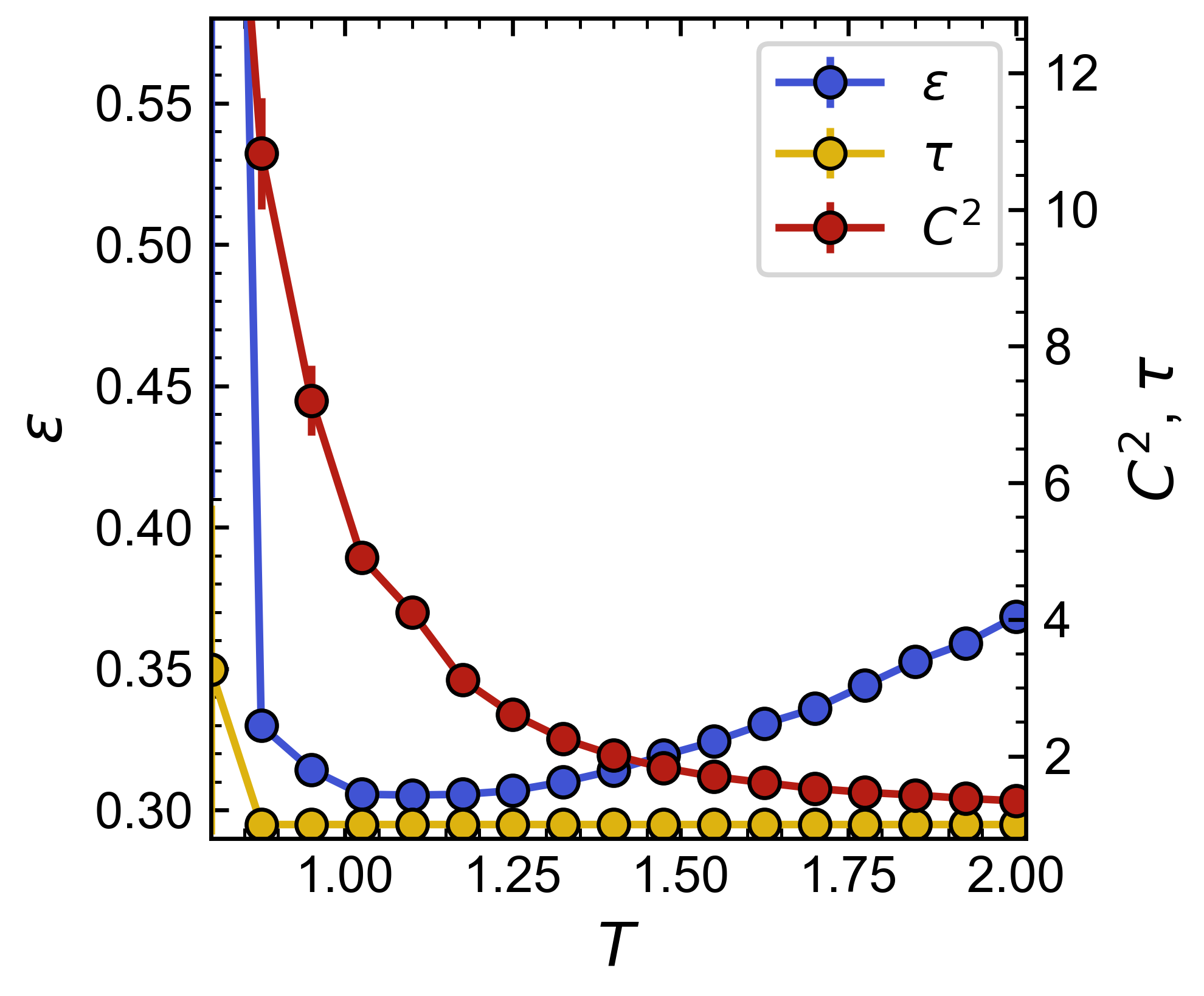}
\caption{Plot of the error $\varepsilon$, auto-correlation time  $\tau$ and correlation measure $C^2$ for varying $T$ at fixed $\mu = 0.1$. Each point is calculated by averaging over MC trajectories of length $B = 10^{4}$ with sampling frequency $10^{3} \times N$ generated from $21$ independent realisations of the SK model at that temperature. Error bars are the standard errors of the observables over these $21$ trajectories. Values $T < 0.8$ are not plotted as $\tau$ began to substantially deviate from $1$ on the approach to the SG regime.}
\label{fig:ObsMuCut}
\end{figure}

The region of minimal error is characterised by iso-contours that run parallel to the phase transition lines, implying a strong dependence of the error on the \textit{distance} from the phase boundary. This is especially clear at low $\mu$, where the $\varepsilon$-contours lie roughly along lines of constant $T$. To understand the relationship between the inference error and the emergence of criticality, we also study a profile of the phase diagram at fixed $\mu=0.1$, away from the ferromagnetic phase. In Fig.~\ref{fig:ObsMuCut} we compare the inference error and $C^2$ as a function of $T$. This confirms that for $N=200$, $\err$ has a flat minimum, centered around a temperate $T_{\text{min}} \approx 1.1$. The shape of the minimum is asymmetric, with $\err$ diverging slowly as $T$ goes from $T_{\text{min}} \to \infty$ and rapidly as $T_{\text{min}} \to 0$. At the minimum, the fluctuations  $C^2(T_{\text{min}})$ are 3-4 times larger than their high-temperature limit. But as $T$ decreases further, the fluctuations continue to increase while the error rapidly diverges. The divergence occurs without a significant increase in the autocorrelation time, indicating that it is not due to poor sampling. The minimum error is therefore within the regime of enhanced critical fluctuations, and occurs close-to but offset-from the phase transition. At the supposed finite size critical temperature the inference fails due to the inherently highly correlated nature of the data from this regime ($\tau$ is large).

\begin{figure}
\includegraphics{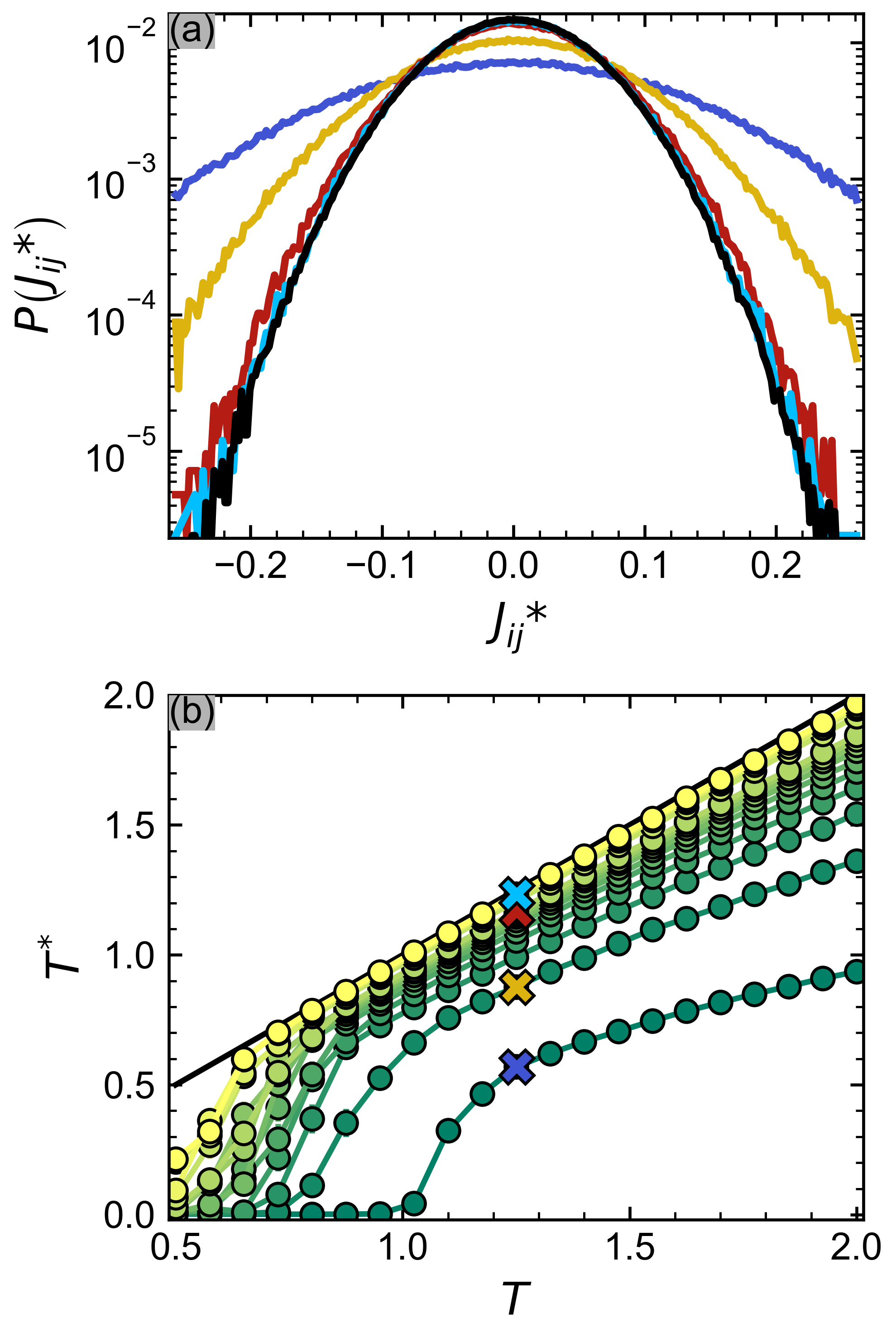}
\caption{
(a) Probability distributions of the inferred couplings from the PLM inference for different varying $B$ from the same state-point $(\mu = 0.1, T = 1.25)$ near the minimum $\err$ in Fig. \ref{fig:ObsMuCut}. Dark-blue, orange, red and light-blue lines correspond to $B= \{ 1,2,10,50 \} \times 10^3$ respectively. The black line shows the ground truth distribution for reference. (b) The inferred temperature as a function of the input temperature for $B = \{ 1,2,3,4,5,6,7,8,9,10,20,30,40, 50 \} \times 10^3$. The darkest green line shows results for the smallest $B$ and the brightest yellow line corresponds to the largest $B$. Coloured crosses indicate temperatures of the respective distributions in (a). $N = 200$, with points and error bars representing the mean and standard error of the PLM temperature estimate from MC trajectories length $B$ of 21 independent model realisations at each $T$.
}
\label{fig:SKDistributions}
\end{figure}


\subsection{Error origin and impact on critical fluctuations}

The inference error $\varepsilon$ is the combination of the error on each individual parameter, dominated by the couplings $J_{ij}$. For highly correlated data we know the PLM inference will fail due to separation. We want to better understand the origin of the error when inference is possible, and to do this we directly inspect the probability distribution of $\bm{J}$ for a selection of sample sizes $B$. Fig. \ref{fig:SKDistributions}(a) shows that the inferred distributions remain symmetric and appear approximately Gaussian, but that they systematically over-estimate the variance, gradually converging to that of the input distribution with increasing $B$. The origin of this spread can be explained in terms of the MLE bias - for logistic regression the parameter estimates are known to be biased away from $0$, i.e. over-estimated \cite{firth1993Biometrika}, which in our case spreads the overall distribution of parameters. Since $T^\ast = 1 / (\sigma^\ast N^{1/2})$, applying PLM to small datasets leads to an inaccurate estimate of the state-point, biased towards a lower temperature. For disordered datasets this corresponds to biasing the model towards the near-critical regime. The question is: how large is this effect?

In Fig.~\ref{fig:SKDistributions}(b) we plot the dependence of the inferred temperature on the input temperature for various $B$. Taking the worse case, $B=1000$, as an example, we find that the inference provides incorrect estimates of the state-point in the entire range of temperatures considered. Even at the optimal conditions, where the inference error is minimal,  $T^\ast\approx 0.4 T^0$ mislabelling the state. Parameter estimates which suffered from separation in Fig.~\ref{fig:SKDistributions}(b) are indicated by low $T^{*} \to 0$, as the anomalously large inferred parameters cause the variance to explode. Collecting more data allows the onset of separation to be delayed, and lower temperature state-points closer to the P-SG transition to be correctly characterised.


We expect that the miss-attributed temperatures will cause the inferred models to exhibit falsely enhanced critical fluctuations. This is because the spin-glass correlations $C^2$ and susceptibility $\chi_{SG}$ increase rapidly as $T$ decreases and one crosses the P-SG transition line. To investigate this, we re-simulate the models inferred via PLM for each $T$ and produce an estimate of the correlations $C^2$ corresponding to the inferred model. This process can be summarized as follows. For every state point:

\begin{itemize}[leftmargin=*]
\itemsep0em 
    \item Produce 21 independent datasets of sample size $B$
    \item Extract 21 PLM models (one per dataset)
    \item Run 6 MC simulations using the PLM estimate for $10^5\times N$ steps, sampled every $10N$ steps ; 
    \item Evaluate $C^2$ over each simulation and average.
\end{itemize}

\begin{figure}
\includegraphics{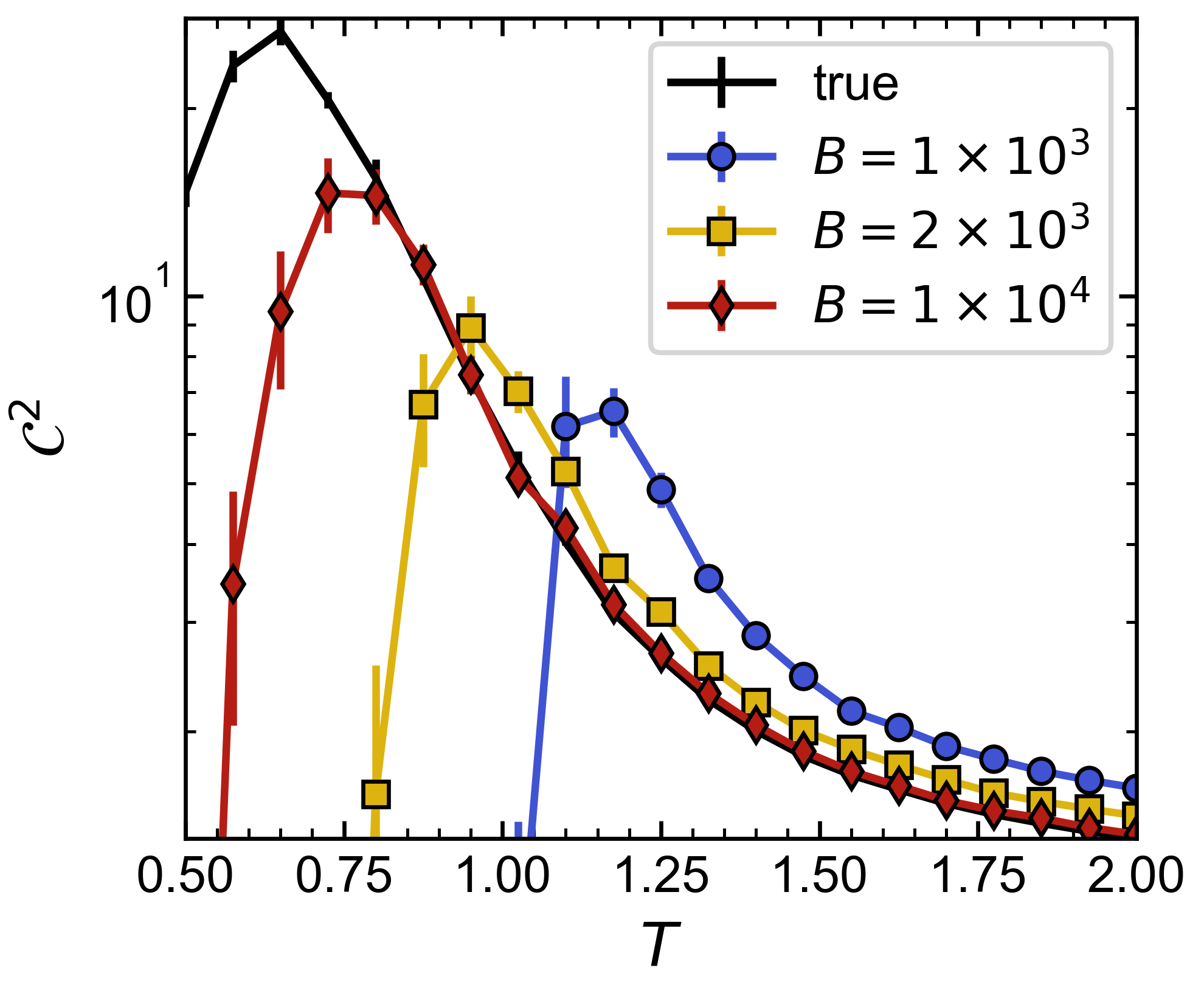}
\caption{Susceptibility measure $C^2$ as a function of input temperature $T$ for three data quality conditions; $B = 1 \times 10^{3}$ (circles), $B = 2 \times 10^{3}$ (squares), $B = 1 \times 10^{4}$ (diamonds). The black line shows the susceptibility as measured from simulations of the true model. Each data-point and error-bar in the black line represents the mean and standard error calculated over $3$ (for each $B$ condition) $\times$ $21 = 63$ independent model realisations.}
\label{fig:SK-C2-impact}
\end{figure}

Fig.~\ref{fig:SK-C2-impact} shows the corresponding results for $C^2$; the black line represents the correlations of the input models, while the colored symbols show those corresponding to the PLM estimates for three different sample sizes. As suspected from the temperature shifts in Fig.~\ref{fig:SKDistributions}, we observe that PLM over-estimates $C^2$ for input models generated at high $T$. With decreasing $T$, $C^2$ reaches a peak value, which gets higher and is located at closer 
 in $T$ to the true $C^2$ peak with increasing $B$. For small sample sizes, the location of the peak corresponds to the onset of separation, as the arbitrarily strongly couplings found for $T<T_{peak}$ fix $C^2 \to 0$. In summary, our numerical experiments on the SK model indicate that PLM on small datasets provides couplings that under-predict $T$ and artificially enhance $C^2$. PLM will therefore miss-attribute finite datasets stemming from the fluctuating P phase to a near-critical state point.



\subsection{Temperature dependence on sample size}

\begin{figure}
\includegraphics{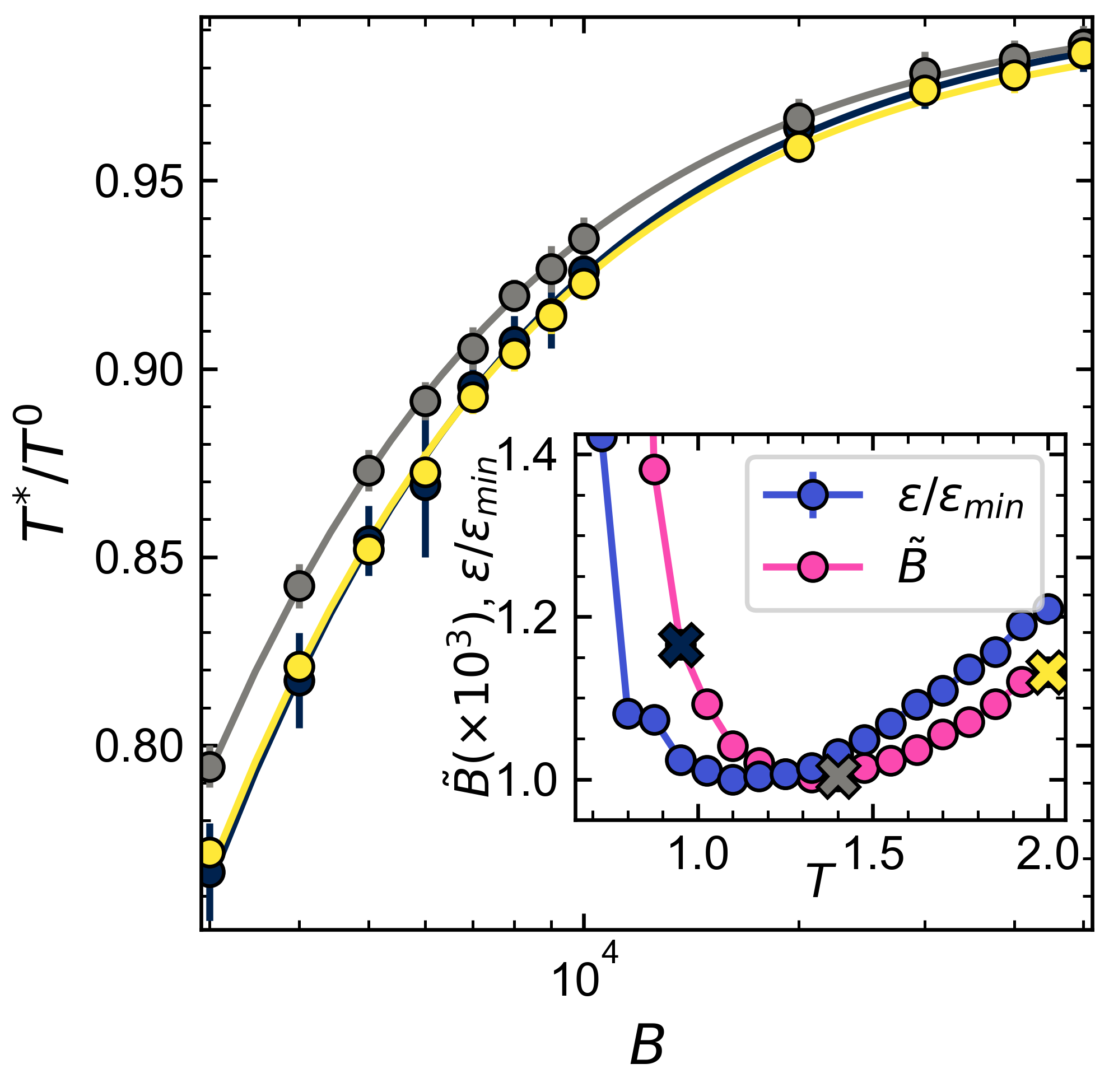}
\caption{$T ^{*} / T ^{0}$ as a function of $B$ for three illustrative input temperatures; (dark-blue) $T = 0.95$, (grey)  $T = 1.4$ and $T = 2$. Points and error bars for each $B$ are means and standard errors obtained by repeating the simulation and inference process for 21 independent input model realisations at each $T$. (Inset) The dependence of the empirical scaling parameter $\tilde{B}(T)$ and the error $\err(T)$ when $B = B_{\text{max}} = 5 \times 10^4$ samples are used for PLM. Crosses show $\tilde{B}$ for the correspondingly coloured $T ^{*} / T ^{0}(B)$ saturation curves in the main body of the figure. Coefficient of determination $R^2 > 0.980$ of the heuristic arc-tan fit (\ref{eq:SatArctan}) for all $\tilde{B}$ plotted.}
\label{fig:SKSaturation}
\end{figure}

In the previous section we qualitatively described that the inferred temperature depends on $B$. Here we quantitatively demonstrate that this effect is governed by the slow $1/B$ convergence of the MLE bias, and that the temperature of the input model sets the learning difficulty of the problem. Fig. \ref{fig:SKSaturation} shows the dependence of the ratio $T^\ast/T^0$ on the sample size for different $T$, where again $\mu=0.1$.  We find that for increasing $B$ the curves follow a saturating behaviour, which can be fitted with high accuracy to the following heuristic model:

\begin{equation}
\Tast = (2 T_{B \to \infty} / \pi) \times \arctan(B / \tilde{B}),
\label{eq:SatArctan}
\end{equation}

where $T_{B \to \infty}$ and $\tilde{B}$ are fitting parameters of the heuristic model. $\tilde{B}$ quantifies the rate of the asymptotic first order bias convergence, and is shown in the inset of Fig. \ref{fig:SKSaturation}, alongside with $\err$. Both share a non-monotonic behaviour indicating a correspondence between the scale $\tilde{B}$ (quantifying the typical sample size to have small deviations in $T^\ast/T^0$) and the average error on the couplings $\err$. The minimum of $\Tilde{B}$ is shifted further into the P phase, to $T \approx 1.4$. Note that for $N=200$, the minimum number of samples is $\tilde{B} \gtrsim 1000$, pointing to the necessity of a minimum of several thousand samples for reliable inference. This highlights a poignant issue for real datasets; the bias dissipation parameter $\Tilde{B}$ is not known a priori, and can vary by orders of magnitude depending on the temperature of the input model. The limit defining a ``small'' dataset therefore depends on the underlying state-point of the data. We motivate the arc-tan fit by noting that for $x=B/\Tilde{B} \ge 1$

\begin{equation}
\arctan(x) = \frac{\pi}{2} - \frac{1}{x} + \mathcal{O}(x^{-3}),
\label{eq:arctan-expansion}
\end{equation}

so that

\begin{equation}
\Tast = T_{B \to \infty} - \frac{\Tilde{B}'}{B} + \mathcal{O}(B^{-3}),
\label{eq:arctan-expansion2}
\end{equation}

follows the same first order linear dependence on $1/B$ as (\ref{eq:MLE-bias}). An equally valid approach would be to plot $T$ as a function of $1/B$ and perform a linear fit to the asymptotic regime as is done in Fig.~\ref{fig:bias-overview}. These results imply that the dependence of the MLE bias prefactors $b_{1, ij}$ on the input model parameters $(\bm{h}^0, \bm{J}^0)$ can, to a good extent, be approximated by a statistical average of the parameter distribution, i.e. $b_{1, ij}(\bm{h}^0, \bm{J}^0) \approx b_{1, ij}(T^0)$.

We have chosen only to present results for $N=200$ here as the analysis of other system sizes lead to identical conclusions. Although there is a sharpening of the phase transition with increasing $N$ the minimum error state-point remains off-set from the transition temperature within the P phase. We note that 
 we find $\Tilde{B} \propto N$, i.e. the amount of data to solve the problem depends linearly on $N$. This dependence arises from the fact that although the full inverse Ising problem deals with estimating $N + N(N-1)/2$ parameters, each individual logistic regression equation only infers $N$ parameters.

In summary, the PLM method on the SK model displays biases that scale as the inverse of the sample size $B^{-1}$. The magnitude of the bias strongly depends on the state-point of the input data. Small sample bias causes the temperature of the inferred model to be under-estimated, falsely enhancing the critical fluctuations exhibited by models inferred from near-critical paramagnetic data. Any PLM model inferred from fluctuating (i.e. dynamically varying) data will thus appear as closer-to-critical than it actually is. In the following, we describe data-driven procedures to mitigate these effects.

\section{Data-driven bias reduction}
\label{sec:correction}

Due to the $1/B$ convergence of the PLM temperature estimate we suspect that methods which remove the first order MLE bias may provide better estimates of the state-point. A range of such methods exist in the literature \cite{kosmidis2014WIREsComput.Stat.}, and can be largely grouped into explicit methods which correct for the bias correction \textit{after} inferring the parameters, e.g. jackknife re-sampling \cite{miller1974Biometrika}, and implicit methods which correct the bias \textit{during} inference via a modification (penalization) of the likelihood function \cite{firth1993Biometrika, heinze2002Stat.Med., zorn2005Polit.Anal., kosmidis2009Biometrika, kosmidis2010Electron.J.Stat.}.

In this section, we propose a new explicit correction to the PLM parameter estimates, which aims to best capture the critical properties of the data by enforcing self-consistency with $C^2$. We benchmark this against an implicit bias correction, Firth's penalized logistic regression \cite{firth1993Biometrika, heinze2002Stat.Med., zorn2005Polit.Anal.}. We will assess both methods based on their ability to estimate $T$ and $C^2$ for a range of input temperatures.

\subsection{Correcting the bias via self-consistent $C^2$}

\begin{figure}
\includegraphics{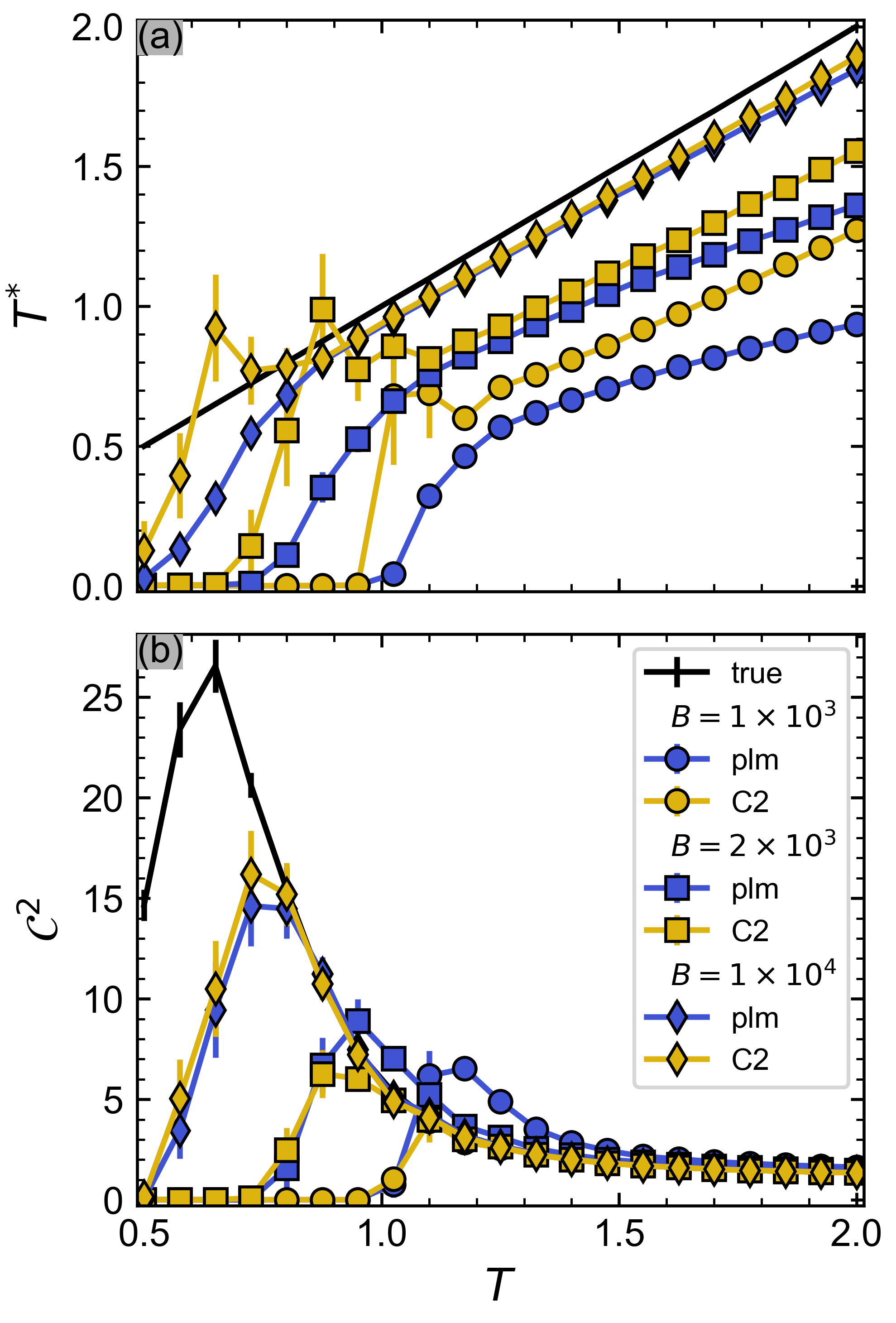}
\caption{The inferred temperature $T^*$ (panel a) and the susceptibility measure $C^2$ (panel b) at different input temperatures $T$ for three data quality conditions; $B = 1 \times 10^{3}$ (circles), $B = 2 \times 10^{3}$ (squares), $B = 1 \times 10^{4}$ (diamonds). Blue lines show observables for the PLM inference,  orange lines show observables after performing the self-consistency (C2) correction. Black lines are the temperature and susceptibility of the input models at each state-point. $N=200$, $\mu=0.1$, with all plotted points again corresponding to means and standard errors over 21 model realisations.}
\label{fig:CorrectionTC2}
\end{figure}

In section \ref{sec:PLMperformance}, we demonstrate that a) the MLE bias of the parameters is captured by a global property of the parameter distribution (the temperature) and b) that the PLM models over-estimate $C^2$. We thus propose to correct the bias by requiring that the inferred models display $C^2$ as close as possible to the one estimated from the input data: in this sense, we aim at inferring models whose fluctuations are self-consistent with those of the input dataset. To do so, we perform a second optimisation \textit{after} estimating the PLM parameters. Again denoting the PLM parameter estimates by $\bm{\theta}^{*}$, we optimize the objective function

\begin{equation}
 \mathcal{L'}(T_{f}) = \left[ C^2 _{\text{input}} -  C^2 _{\text{MC}}(T_{f}) \right] ^2,
\label{eq:Correction}
\end{equation}

where $C^2 _{\text{input}}$ is $C^2$ measured as from the input dataset and $C^2 _{\text{MC}}(T_{f})$ is calculated from MC simulations of a re-scaled PLM model with parameters $\bm{\theta}_{T_f}= \boldsymbol{\theta}^* / T_f$. The re-scaling parameter $T_f > 0 $ acts as a \textit{fictitious temperature} which can shift the state-point of the inferred model. We denote optimal value of $T_f$ by $T_f^\dagger$, with the corresponding corrected parameter estimates being $\bm{\theta}^\dagger$. $C^2$ has a strong dependence on $B$, so we match the amount of data in the input and in the MC simulations $B_{\text{data}} = B_{\text{MC}}$. In practice we calculate $C^2 _{\text{MC}}$ for $6$ independent MC simulations of length $B_\text{data}$ for every $T_f$, and then feed the average over these $6$ runs into (\ref{eq:Correction}).

The results of this corrective procedure are shown in Fig.~\ref{fig:CorrectionTC2}. The correction significantly improves the reconstructed temperature and, by design, perfectly matches the fluctuations of $C^2$ when separation does not occur. The improvement to the $T^{*}$ prediction is particularly pronounced for small datasets and at high $T$. At low $T$, where separation occurs, the corrective optimization fails to converge and $C^2 _{\text{input}} \neq C^2(T_f ^\dagger)$. This highlights a pitfall of explicit methods; they inherit any instabilities of the original MLE, such as those that lead to separation in logistic regression \cite{zorn2005Polit.Anal., kosmidis2014WIREsComput.Stat.}. We note that the improvement itself can also be used as a score on the reliability of the PLM estimates and as an indication of the necessity of more data, with $T_f \to 1$ as $B \to \infty$.

\subsection{Firth's penalized logistic regression}

\begin{figure}
\includegraphics[width=\linewidth]{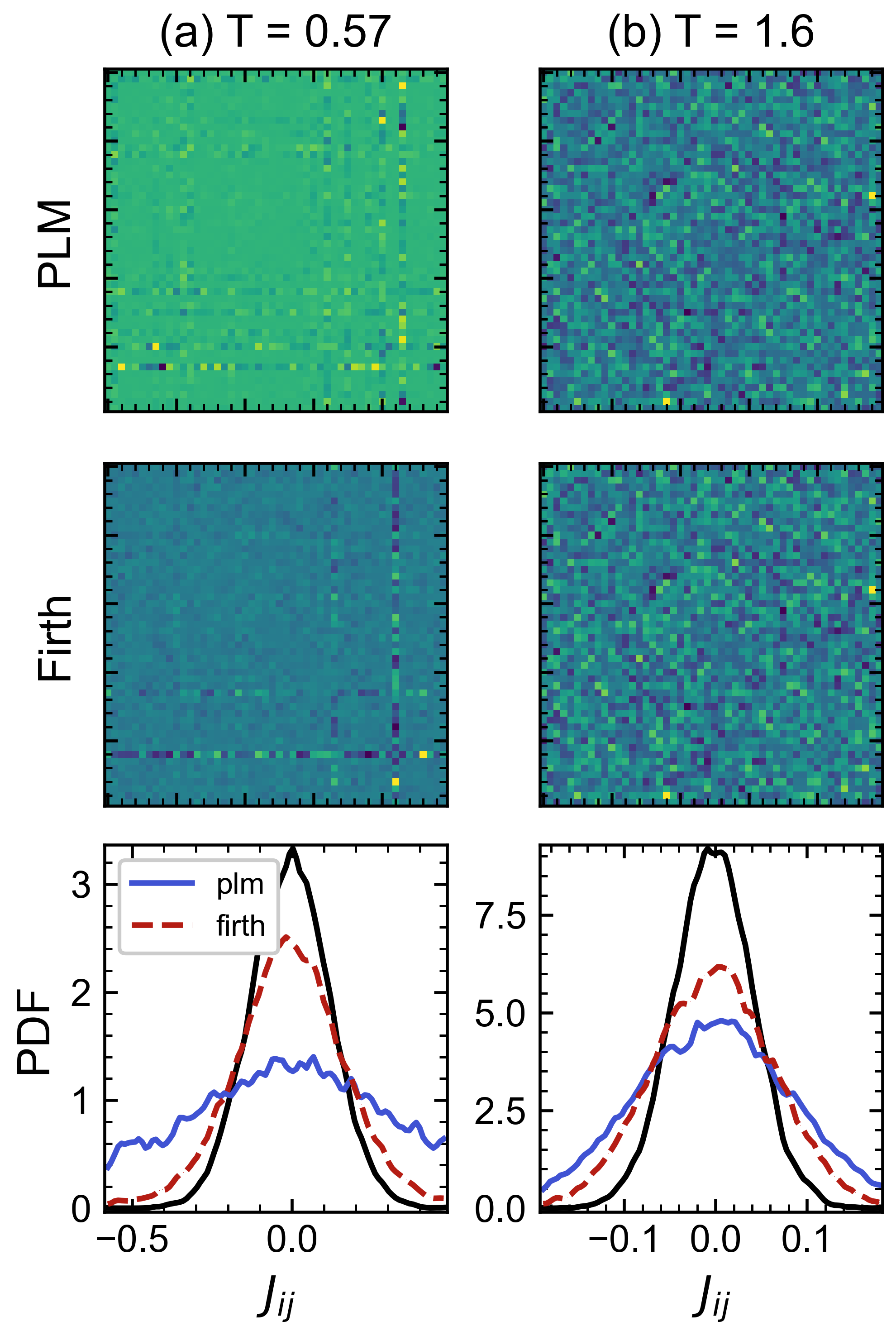}
\caption{Parameter matrices $\bm{\theta}$ for a subset of $50$ spins for un-penalized PLM (top) and Firth's penalized PLM (middle) at two different temperatures. (Bottom) probability density functions (PDFs) of the corresponding inferred parameters along with the true parameter PDF (black line). Firth's correction controls the inference at low $T$, leading to finite parameters and a PDF that more closely matches the true input model. At high $T$ firth's correction shifts the inferred temperature towards the true temperature (by reducing the spread of the parameter PDF). $B=10^{3}$, $N=200$ and $\mu=0.1$.}
\label{fig:firth-plm-mod-dist}
\end{figure}

One may also remove the first order bias implicitly through Firth's penalized likelihood maximisation \cite{firth1993Biometrika}. In our notation, this corresponds to maximising the penalized log-likelihood $\mathcal{L}'_r$:

\begin{equation}
    \mathcal{L}'_r(h_r, \bm{J_r} | \{ \bm{s} \}_B) = \mathcal{L}_r(h_r, \bm{J_r} | \{ \bm{s} \}_B) +
    0.5 \ln | F(h_r, \bm{J_r}) |,
\label{eq:firth}
\end{equation}

where $| F(h_r, \bm{J_r}) |$ is the determinant of the Fisher information matrix for each row of parameters $r$. Full details of this method will not be given here, but see \cite{firth1993Biometrika, zorn2005Polit.Anal., kosmidis2014WIREsComput.Stat.} for appropriate overviews. We implement this computationally by modifying the Python code available at \url{https://github.com/jzluo/firthlogist}. The corrective term of eq.~\ref{eq:firth} tends to $0$ as $B \to \infty$, returning the un-penalised likelihood. For small $B$ the penalty compensates the $\mathcal{O}(B^{-1})$ bias, and is known to control separation in logistic regression \cite{heinze2002Stat.Med., zorn2005Polit.Anal.}.

In Fig.~\ref{fig:firth-plm-mod-dist} we demonstrate the effect of the penalty on the inferred parameters at a low $T=0.57$ and a high $T=1.6$ state-point. At low $T$ separation causes the un-penalized PLM parameters associated with specific $s_{i}$ to diverge, with $\max(\theta_{ij}^{\text{PLM}}) \approx 400$, leading to a non-Gaussian highly spread PDF. In our language models with these very strong couplings corresponding to zero temperature. Firth's penalty controls this effect, and although we still see large parameter estimates for the same $s_{i}$, these are orders of magnitude smaller with $\max(\theta_{ij}^{\text{Firth}}) \approx 4$. Firth's correction reduces the spread of the inferred distribution and largely captures the Gaussian nature of the input coupling distribution, even at low $T$. It therefore provides better $T^{*}$ estimates than un-penalized PLM.

\subsection{Comparing methods}

\begin{figure}
\includegraphics[width=\linewidth]{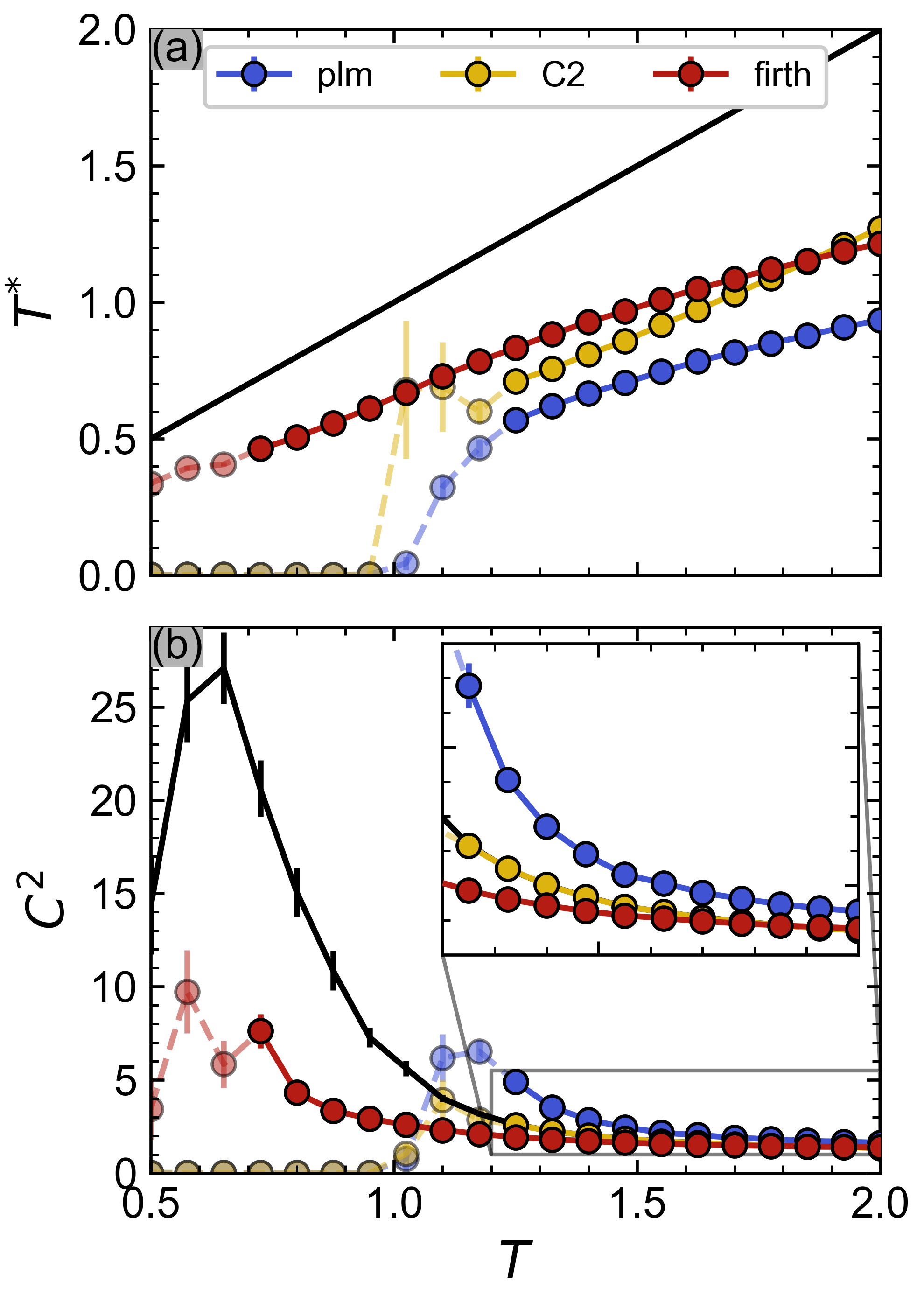}
\caption{Comparison of the corrective methods ability to capture the criticality relevant observables $T^*$ and $C^2$ for a small sample size $B=10^{3}$. Transparent points indicate $T$ where separation occurred. Firth's correction gives reasonable estimates for $T^*$ at much lower $T$ than PLM, highlighting this methods ability to control separation. In contrast to normal PLM, Firth's correction under-estimates $C^2$ for all $T$. At high $T$ Firth's correction and the C2 correction perform similarly. Each point represents average over 21 independent datasets at that $T$.}
\label{fig:firth-T-C2}
\end{figure}

In Fig.~\ref{fig:firth-T-C2} we compare the performance of the self-consistency (C2) correction and Firth's (firth) correction for the smallest dataset, $B=1000$, where bias effects are most extreme. We again generate data from $21$ independent model realisations for each $T$ and then apply each inference method to each dataset separately. We naively assess if separation has occurred by checking if $|\bm{\theta}^{*}| _{max} > \mu^{*} + 10 \sigma^{*}$, where $|\bm{\theta}^{*}|$ is the absolute value of the inferred parameters from each inference scheme. Any input $T$ where \textit{a single} inferred model satisfied the separation condition (i.e. had ``anomanlously'' large parameters) is indicated by the transparent points in Fig.~\ref{fig:firth-T-C2}. We note that this is a conservative definition; if even a single inference fails we characterise the whole state-point as separation prone. We see that (according to our definition) the onset of separation is delayed to much lower $T$ when using Firth's penalty, and that Firth's penalised logistic regression predicts non-zero $T^{*}$ for all $T$. At high $T$ both corrections perform similarly well in improving the estimated $T^{*}$, although the dependence $T^{*}(T)$ appears to scale more favourably using the C2 correction as $T$ increases.

Fig.~\ref{fig:firth-T-C2}(b) shows how well each method reproduces the correlations of the input data. We observe an interesting trade off; although Firth's correction provides better estimates for $T$, $C^2$ of the corresponding models is systematically under-predicted. Firth's correction thus fails to capture the correlations of the data. We note that this contrasts with the un-penalised logistic regression results for PLM, which instead over-estimate $C^2$. The main advantage of Firth's correction over the C2 correction, therefore, appears that lower temperature sate-points can be estimated. 

This naturally raises a question: can we apply the C2 correction to the models inferred using Firth's penalization and improve the estimate of both $T$ and $C^2$? The answer is negative: unsurprisingly, applying the C2 correction to the penalized parameter estimates increases $C^2$ at the cost of lowering $T^{*}$, and ultimately leads to the same estimates as simply applying the C2 correction to the un-penalized PLM model. We summarise these findings in Table~\ref{table:infmethod-comparison}. There thus appears to be an inherent trade off between capturing the temperature and the correlations of a dataset. When deciding which method is ``best'' one must therefore decide which property is most important to encode correctly.




\begin{table}
\centering
\begin{ruledtabular}
\begin{tabular}{l  c  c } 
 Method & $T^{*}$ $\%$ error & $C^2$ $\%$ error \\ 
 \hline
 \hline
 PLM                    & -6.9  & 3.4(8)  \\ 
 PLM $\rightarrow$ C2   & -6.0  & 0.2(9) \\
 Firth                  & -4.0  & -7.7(5) \\
 Firth $\rightarrow$ C2 & -6.1  & -0.1(8) \\
\end{tabular}
\end{ruledtabular}
\caption{Percentage errors of inferred estimates $T^{*}$ and $C^2$ for a single model realisation at $T=1.025$, $\mu = 0.1$, with $B=10^{4}$ using a range of inference schemes. Firth's correction provides the best estimate of the temperature but the worst estimate of the critical fluctuations. We demonstrate an implicit trade off between correctly inferring $T$ and $C^2$. Applying the C2 correction either to the PLM model or to the Firth corrected model produces the same $T^{*}$ and $C^2$ pair. The temperature is under-estimated, irrespective of the inference scheme used.}
\label{table:infmethod-comparison}
\end{table}


\subsection{Implications for inference around criticality}

So far we have shown that small-sample biases influence the determination of the state of Ising models inferred using PLM. The problem is that what constitutes a ``small'' sample size itself depends on the state-point (and topology) the true model that generated the data. Studies claiming criticality in Ising models inferred using PLM thus need to control for bias, for example through sub-sampling their data and performing a similar analysis as in Fig.~\ref{fig:SKSaturation}. They should also consider that the PLM model they infer from any dataset which is dynamic (i.e. fluctuates) will be biased \textit{towards} the critical point and exhibits enhanced critical fluctuations when simulated. In such cases, the C2 correction may be applied to re-scale the couplings and match the empirical correlations. This will also shift the inferred temperature towards the true temperature. We note, however, that the C2 corrected temperature remains systematically smaller than the true temperature, and should only be considered as a lower-bound estimate of the true temperature. It may be appropriate to use Firth's implicitly corrected penalized logistic regression if separation is found to occur. This correction will provide reasonable parameter estimates even for low $T$ state points, but it should be noted the fluctuations displayed by these models are not representative of the data, which is important when considering near-critical phenomena.

\section{Case study: Criticality of an fMRI dataset}
\label{sec:neuro-case-study}

In this section we demonstrate the effect of the bias in a real setting, by analysing a human brain imaging dataset obtained from a \textit{functional magnetic resonance imaging} (fMRI) study. Brain imaging datasets are particularly interesting from the perspective of criticality as a range of advantageous computational properties have been attributed to systems operating in the near-critical state \cite{wilting2019CurrentOpinioninNeurobiology, kinouchi2006NaturePhys, cavagna2010ProceedingsoftheNationalAcademyofSciences, tagliazucchi2012Front.Physiol., ramo2006JournalofTheoreticalBiology, deco2012JournalofNeuroscience, bertschinger2004NeuralComputation, legenstein2007NeuralNetworks, haykin2006, beggs2003J.Neurosci., beggs2004JournalofNeuroscience, plenz2012Eur.Phys.J.Spec.Top., yu2013Front.Syst.Neurosci., marinazzo2014PLoSONE, bialek2018Rep.Prog.Phys.}. These have given rise to the idea that the brain is tuned towards a critical point, and a range of experimental results support this \textit{critical brain hypothesis} \cite{beggs2003J.Neurosci., beggs2004JournalofNeuroscience, haldeman2005Phys.Rev.Lett., beggs2008PhilosophicalTransactionsoftheRoyalSocietyA:MathematicalPhysicalandEngineeringSciences, friedman2012Phys.Rev.Lett., beggs2012Front.Physiol., plenz2012Eur.Phys.J.Spec.Top., munoz2018Rev.Mod.Phys., obyrne2022TrendsinNeurosciences}. PLM was recently used to contribute to this body of work, correlating IQ to the spin-glass susceptibility \cite{ezaki2020CommunBiol}, and inverse Ising inference in general has previously been used to map different cognitive states to different disordered spin models \cite{watanabe2014NatCommun}.


We consider data from a single-participant resting-state fMRI study, the full experimental details of which can be found in \cite{kajimura2020SciRep}. Imaging sessions were carried on separate days under two different conditions: those where the participant practiced mindfulness meditation (MM) before undergoing imaging, and those where he did not (noMM). During each session $B=236$ samples were collected from $N=399$ regions of interest (ROIs) within the brain. We will consider each ROI as a spin $s_i$ and perform inverse Ising inference using PLM. Note that the trajectories for each ROI, $s_{i}(t)$, obtained from the pre-processing in \cite{kajimura2020SciRep} are continuous. We therefore binarised the data by removing the average from the signal and setting $s_{i}(t) < 0 = -1$ and any $s_{i}(t) \ge 0 = +1$. In total $B_{\text{noMM}} = 40 \times 236 = 9440$ samples were collected for the noMM condition and $B_{\text{MM}} = 18 \times 236 = 4248$ for the MM condition. We investigate whether PLM inference leads to the identification of a close-to-critical state, and if this state is affected by the biological condition (i.e. practicing of mindfulness meditation). Moreover, we study if the inference biases identified here significantly impact our conclusions.


\begin{figure}
\includegraphics{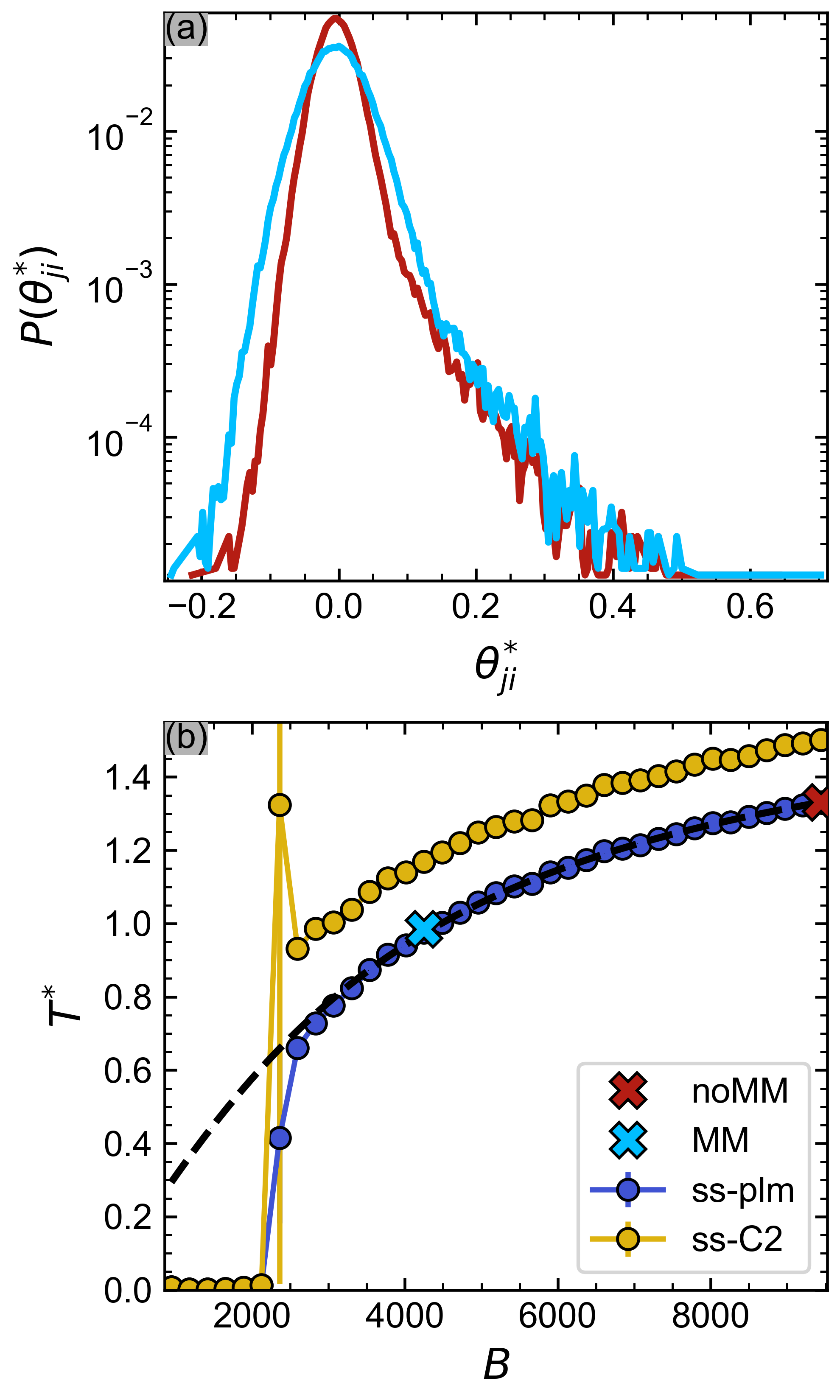}
\caption{(a) Inferred parameter distributions for the noMM (red) and MM (blue) conditions. The bias causes the MM distribution to appear more spread. (b) Sub-sampling analysis of the noMM dataset. (blue) PLM temperature estimates for each number of sub-samples, and (orange) the corresponding temperature of the self-consistency corrected model. The fitted empirical model (\ref{eq:SatArctan}) is shown by the dashed black line. Coloured crosses show the temperatures corresponding to the distributions in (a).}
\label{fig:kajimura-dist-and-subsample}
\end{figure}

The distributions of the PLM parameters obtained from the \textit{full} datasets are shown in Fig.~\ref{fig:kajimura-dist-and-subsample}(a). First, as opposed to the SK model, the distributions are skewed, with a long tail at positive values of the couplings. Second, the MM condition corresponds to a larger variance in the couplings than the noMM one. The immediate consequence is that the mapped temperatures of the two full datasets are different, $T^{*} _{\text{full-MM}} = 0.98$ and $T^{*} _{\text{full-noMM}} = 1.33$.

However, the two sample sizes are $B_{\text{noMM}} \approx 2 B_{\text{MM}}$. Can this lead to a significant statistical effect in the estimation of the corresponding state points? Fig.~\ref{fig:kajimura-dist-and-subsample}(b)  provides an affirmative and quantitative answer: as we sub-sample (ss) the noMM data we find that the PLM temperature estimate decreases with reducing $B$, and when $B_{\text{ss-noMM}} = B_{\text{MM}}$, meets the same estimated temperature of the MM data. Hence, for the data considered here, there is no significant difference between the MM and the noMM data when the statistical bias is taken into account. We further find that below some critical value $B_c \approx 2000$ separation occurs leading to the failure of the inference.

To refine the estimate of the noMM state-point, we can apply the self-consistency correction (yellow circles) and fit the empirical saturation function, Eq.~\ref{eq:SatArctan}, to the PLM temperature estimates. Fitting data with  $B \ge 2500$, we find $T_{B \to \infty} = 1.72$ and $\tilde{B} = 3465$. The available datasets are with $B_{\rm MM}\approx 1.2\tilde{B}$ and $B_{\rm MM}\approx 2.7\tilde{B}$, indicating that the small sample size bias strongly impacts our conclusions here.

\begin{figure}
\includegraphics{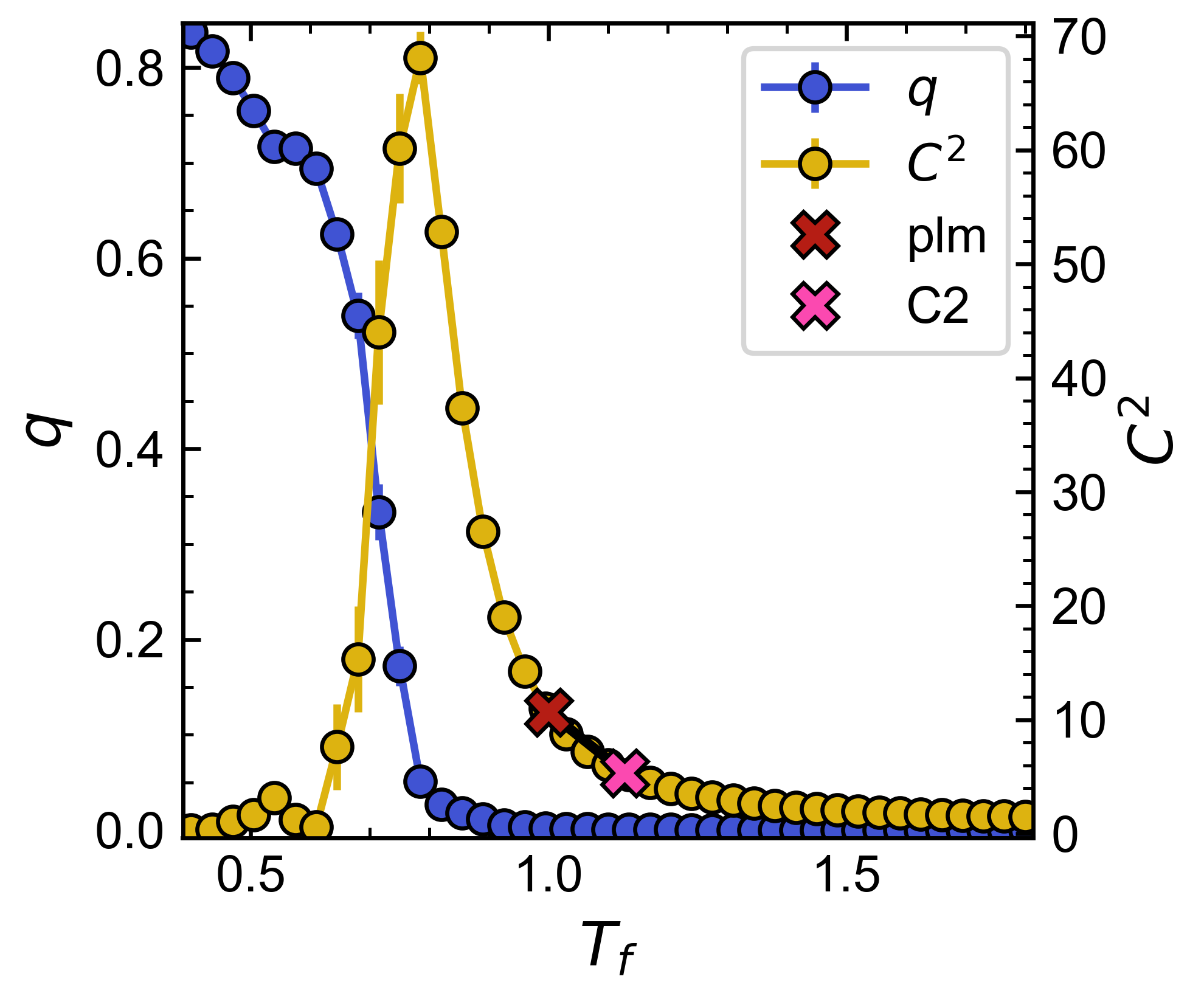}
\caption{Spin glass order parameter $q$ and susceptibility $C^2$ as functions of the fictitious temperature $T_f$ for the model inferred from the no-MM condition, $\bm{\theta}^{*}_{\text{noMM}}$. The ferromagnetic order parameter $m$ was also measured, and found to be $0$ throughout. This sweep therefore characterises a transition from a low-temperature spin-glass to a high temperature paramagnetic phase. Points and error bars correspond to means and standard errors of 60 independent MC simulations with $B=10^4$ samples. Red and pink crosses correspond respectively to the PLM and self-consistency corrected models of the noMM condition.}
\label{fig:kajimura-statepoint}
\end{figure}

To contextualise the meaning of the inferred temperatures in Fig.~\ref{fig:kajimura-dist-and-subsample}(b) we again introduce the fictitious temperature $T_f$ and perform MC simulations of $\frac{1}{T_f}\boldsymbol{\theta}^{*}_{\text{noMM}}$ for a range of $T_f$, see Fig.~\ref{fig:kajimura-statepoint}. A peak of $C^2$ at $T_f = 1$ would mean that the PLM model is situated exactly at the critical point. We instead find the peak at $T_f = T_c = 0.78$ and so $\boldsymbol{\theta}^{*}_{\text{noMM}}$ is a paramagnetic state-point above the transition, albeit still with substantial critical fluctuations $C^2(T_f = 1) \approx 0.15C^2_{\text{max}}$. The self-consistency correction shifts the model further from $T_c$. Hence, the MM and noMM conditions appear to be at best-near critical if not paramagnetic.

Summarising these results, initial analysis of the two conditions would lead to the conclusions that a) practicing mindfulness meditation changes the state-point of the brain, and b) that the noMM condition represents a near-critical paramagnetic state-point. Carefully accounting for the bias instead reveals that both datasets more likely originate from the same state-point, and the C2 corrected temperature estimate shows that, as a lower bound, the true state-point of the data lies far from the transition in the paramagnetic phase. We therefore find no evidence to suggest that the resting state fluctuations in this imaging study correspond to a critical brain state.


\section{Conclusions}

In this work, we have studied the importance of small sample size biases in the pseudo-likelihood maximisation (PLM) approach to inverse Ising inference. Although PLM is exact in the limit of large sample size, we show that this condition is often unlikely to be achieved in real world datasets. We demonstrate that estimates of important physical quantities (such as the temperature) which define the state of the inferred model depend linearly on $1/B$, similarly to the standard bias of maximum likelihood estimators.

We present a detailed study of the fully connected SK model for $N=200$. The above biases cause models inferred from paramagnetic datasets to exhibit enhanced critical fluctuations, with this effect worsening with decreasing $B$. Paramagnetic data is therefore miss-classified as near-critical for finite $B$, i.e. PLM under-estimates the distance from criticality. The inference error is minimized close-to but off-set from the phase transition in the paramagnetic regime. The development of strong correlations on the approach to the critical point means that PLM fails due to separation at lower $T$. We note that, although information theoretic arguments suggest otherwise \cite{mastromatteo2011J.Stat.Mech.}, for small or intermediate $B$ the regime of failure occurs \textit{before} the finite size critical temperature $T_c$ is reached.

We describe data-driven approaches to mitigate these effects. The self-consistent correction we propose improves the temperature estimate $T^{*}$ while matching the critical fluctuations of the dataset. It performs well when a PLM solution can be found, i.e. when separation does not occur, and provides the best improvement to $T^{*}$ at high $T$. For low $T$ (or equivalently for small $B$) Firth's penalized logistic regression may be used to estimate the state-point when standard PLM fails. Although this produces $T^{*}$ closer to $T$, we caution that the critical fluctuations inferred using Firth's correction are not representative of the data. These models thus fail to capture an essential property of the system. Both the self-consistency correction and Firth's correction provide biased estimates of the temperature, with $T^{*} \to T^{0}$ from below as $B \to \infty$. The estimated temperatures $T^{*}$ should therefore be considered as lower bounds on the true temperature of the dataset. In contrast to other regularization techniques, neither correction requires a hyperparameter to be tuned.

We also show that the bias profoundly impacts the estimation of criticality in a real finite $B$ dataset from neuroscience. Not accounting for small sample size effects causes state-points to be incorrectly classified. For fluctuating, i.e. dynamically varying data, this corresponds to inferring models which are falsely tuned towards the critical point. Applying the self-consistency correction to this dataset allows us counteract this and establish that, as a lower bound, the data is paramagnetic. The above results leads us to conclude that any PLM study claiming criticality in a real dataset must carry out a proper analysis of the dependence of the inference on $B$, e.g. through the sub-sampling scheme we describe. Otherwise small sample size biases cannot be ruled out as the primary cause of the criticality of the inferred model. This is especially important as we have shown that the bias prefactors, e.g. $\Tilde{B}$, setting the learning difficulty of the model are functions of the state (and also topology \cite{lokhov2018Sci.Adv.}), and that one thus cannot establish a priori what constitutes a ``small'' sample size.

With a view to the future, experimental evidence for criticality appears to be ripe across biological systems \cite{munoz2018Rev.Mod.Phys.}. Maximum entropy approaches, such as the PLM solution for inverse Ising inference, provide an attractive route 
with which to investigate these, by allowing the criticality of the data to be assessed in the framework of statistical physics. We show here that any such study must make careful consideration of small sample size biases before claiming a system to be critical. This is especially important as the distance to criticality is becoming an increasingly relevant biological variable \cite{obyrne2022TrendsinNeurosciences} and, in neuroscience for instance, is starting to be considered as a guiding route for clinical work \cite{zimmern2020Front.NeuralCircuits}.

\textit{Code Availability -} The code used for this study is available as a Python package at \url{https://github.com/maxkloucek/pyplm}.

\section{Acknowledgments}

This work was supported by the EPSRC Centre for Doctoral Training in Functional Materials: The Bristol Centre for Functional Nanomaterials (BCFN) with grant code EP/L016648/1. N.M. acknowledges support from the Japan Science and Technology Agency (JST) Moonshot R$\&$D (under Grant No. JPMJMS2021).

\end{document}